%% file: manuscript.tex
\documentclass[submission, Phys]{SciPost}
\usepackage{amsmath,cases}
\usepackage[utf8]{inputenc}
\usepackage[T1]{fontenc}
\usepackage{lmodern}
\usepackage{graphicx}
\usepackage{amssymb}
\usepackage{gensymb}
\usepackage{bm}
\usepackage{tabularx}
\usepackage{mathtools}
\usepackage{microtype}
\usepackage{siunitx}
\usepackage[version=4]{mhchem}
\usepackage{bbm}
\usepackage[justification = raggedright]{caption}
\usepackage{braket}
\usepackage{mathtools}
\usepackage[normalem]{ulem}
\usepackage{layouts}
\usepackage{float}
\usepackage{textgreek}
\usepackage{orcidlink}
\usepackage{multirow}
\usepackage{amsmath}
\usepackage{amsfonts}
\usepackage{todonotes}

\hypersetup{colorlinks}

\newcommand{\comment}[1]{\stepcounter{CommentNumber}\belowpdfbookmark{#1}{\arabic{CommentNumber}}}
\newcommand{\ZZ}{\mathbb{Z}}

\newcommand{\vk}{\mathbf{k}}

\newcommand{\gr}[1]{{\color{teal}#1} }
\newcommand{\bl}[1]{ {\color{blue}#1} }
\newcounter{CommentNumber}

\begin{document}

\begin{center}{\Large \textbf{
      Symmetric approximant formalism for statistical topological matter
    }}\end{center}

\begin{center}
  R.~Johanna~Zijderveld\textsuperscript{1$\star$}\orcidlink{0009-0007-6281-5052},
  Adam~Yanis~Chaou\textsuperscript{2,3${}^\S$}\orcidlink{0000-0002-9926-4633},
  Isidora~Araya~Day\textsuperscript{1,3,4$\dagger$}\orcidlink{0000-0002-2948-4198},
  Anton~R.~Akhmerov\textsuperscript{1$\ddagger$}\orcidlink{0000-0001-8031-1340}
\end{center}

\begin{center}
  \small
  \textbf{1} Kavli Institute of Nanoscience, Delft University of Technology, 2600 GA Delft, The Netherlands \\
  \textbf{2} Freie Universität Berlin, Dahlem Center for Complex Quantum Systems and Fachbereich Physik, Arnimallee 14, 14195 Berlin \\
  \textbf{3} Donostia International Physics Center (DIPC), Paseo Manuel de Lardiz\'{a}bal 4, 20018, Donostia-San Sebasti\'{a}n, Spain \\
  \textbf{4} QuTech, Delft University of Technology, Delft 2600 GA, The Netherlands \\
  ${}^\star${\small \sf johanna@zijderveld.de}
  ${}^\S${\small \sf a.chaou@fu-berlin.de}
  ${}^\dagger${\small \sf isidora@araya.day}
  ${}^\ddagger${\small \sf approximant@antonakhmerov.org}
\end{center}

\begin{center}
  August 07, 2026
\end{center}

\section*{Abstract}
\textbf{
The standard approach to characterizing topological matter---computing topological invariants---fails when the symmetry protecting the topological phase is preserved only on average in a disordered system.
Because topological invariants rely on enforcing the symmetry exactly, they can overcount phases by incorrectly identifying certain non-robust features as robust.
Moreover, in intrinsic statistical topological insulators, enforcing the symmetry exactly is guaranteed to destroy the topological phase.
We define a mapping that addresses both issues and provides a unified framework for characterizing disordered topological matter.
}
\section{Introduction}

\comment{Because topology deals with quantization, it is in natural tension with statistical properties.}
Macroscopic observables in condensed matter physics emerge as averages over local quantities, and are therefore inherently subject to statistical fluctuations.
The distinctive feature of topological phases of matter is the existence of quantized observables---the topological invariants---that change discontinuously in the thermodynamic limit as the system undergoes a phase transition.
This quantization creates a tension with statistical fluctuations in disordered systems whose protecting symmetry is only respected on average by an ensemble.
The standard way to understand topology is to group Hamiltonians into equivalence classes that cannot be continuously deformed into each other without breaking the symmetry.
This picture fails for disordered ensembles with average symmetries because such symmetries constrain the probability distribution over Hamiltonians rather than each realization.
Individual disorder realizations therefore break the symmetry and cannot be separated into equivalence classes.

\comment{Statistical topological matter possesses unique phenomenology.}
The symmetry properties of disordered ensembles stabilize topological phases.
Using weak topological insulators as a first example~\cite{Ringel2012, Mong2012, Fu2012}, prior work showed that the average symmetry of the disordered ensemble can pin the surface of the individual realizations to a topological transition point between two gapped phases.
As a result, the surface states avoid Anderson localization, giving rise to a statistical topological phase---an ensemble of Hamiltonians in the same symmetry class where an additional ensemble symmetry protects delocalized boundary states~\cite{Fulga2014}.
For example, the surface of a 3D topological insulator with a random magnetization realizes the quantum Hall critical point~\cite{Nomura2008}, while a time-reversal-symmetric disordered surface is a metal whose conductance grows logarithmically with system size~\cite{Bardarson2007,Nomura2007}.
The challenge of describing statistical topological phases, however, extends beyond their different surface phenomenology.
In multiple symmetry classes, disorder that respects a symmetry only on average results in fewer distinct topological phases than in the clean limit~\cite{Fulga2014,Chaou2025}.
Strikingly, the opposite phenomenon was recently discovered: disorder enables topological phases forbidden in clean systems.
Reference~\cite{Chen2025} introduced intrinsic statistical topological insulators (ISTIs) by demonstrating the existence of a topological phase in a 3D system with an anomalous magnetic rotation symmetry.
Disordered phases therefore require a fundamentally different topological invariant than those used in translationally invariant systems.

\comment{Known approaches rely on the group-theoretic classification, qualitative arguments, and thermodynamic responses.}
The standard method for analyzing topological phases in the presence of disorder is to use group-theoretic arguments to assess the stability of their anomalous boundary states~\cite{Ringel2012,Fulga2014,geier2018,Chaou2025}.
Identifying whether a specific disordered phase is realized then relies either on probing its boundary properties and demonstrating the absence of localization, or on measuring thermodynamic responses that are quantized in the topological phase~\cite{Chen2025}.
While these approaches are sufficient to probe and classify statistical topological phases, they leave an open question: is it always possible to define an invariant for phases that preserve a symmetry only on average?

\begin{figure}[bth]
  \centering
  \includegraphics[width=0.8\columnwidth]{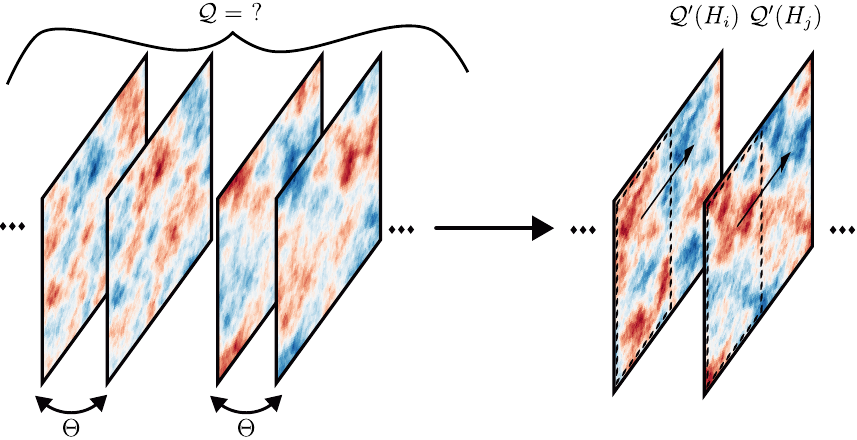}
  \caption{
    \label{fig:formalism}
    An illustration of the symmetric approximant formalism, whose goal is to define a topological invariant $\mathcal{Q}$
    for disordered ensembles with average symmetries.
    Left: A disordered ensemble with an average time-reversal symmetry $\Theta$.
    Every realization $H_i$ breaks time-reversal symmetry but has an equally likely time-reversed partner $\Theta H_i \Theta^{-1}$ in the ensemble.
    Right: The corresponding symmetric approximant ensemble where each realization $H_i$ preserves an exact magnetic translation symmetry $\Theta T_{x, L/2}$ and has a well-defined topological invariant $\mathcal{Q}'(H_i)$.
    The arrow between the ensembles indicates the mapping from the original ensemble to the symmetric approximant ensemble.
  }
\end{figure}

\comment{We demonstrate that a symmetry approximant captures topology in many cases.}
In this article, we show how to define, for a broad class of statistical topological phases, a \emph{symmetric approximant} ensemble: a proxy for the original ensemble that captures the topology of a disordered phase, as depicted schematically in Fig.~\ref{fig:formalism}.
In addition to having the exact symmetries of the original ensemble, the elements of the approximant ensemble respect a subgroup of the average symmetry of the original ensemble.
Because each of the elements of the approximant ensemble respects an additional exact symmetry, we may apply the conventional topological invariants of that symmetry class to the entire approximant ensemble.
The goal is therefore to choose the symmetric approximant such that if the approximant is delocalized, so is the original disordered ensemble, thereby ensuring that both share the same topological phase transitions.

\comment{We construct the formalism in steps that address individual challenges of statistical topology.}
The structure of this manuscript is as follows.
In Sec.~\ref{sec:magnetictranslation} we introduce the mapping using the examples of weak and magnetic statistical topological insulators.
We continue in Sec.~\ref{sec:indistinguishability} by demonstrating how to construct the mapping and the corresponding symmetric approximant for a statistical phase protected by reflection symmetry.
In Sec.~\ref{sec:ISTI} we demonstrate that our procedure naturally supports characterizing intrinsic statistical phases upon breaking translation symmetry.
We generalize our findings in Sec.~\ref{sec:general}, where we categorize the types of statistical phases our procedure supports and identify where it fails.
We conclude in Sec.~\ref{sec:conclusion} by summarizing our results and discussing future directions.

\section{The procedure and its motivating example} \label{sec:magnetictranslation}
\subsection{Enforcing translation symmetry of a supercell}

\begin{figure}[bth]
  \centering
  \includegraphics[width=0.9\columnwidth]{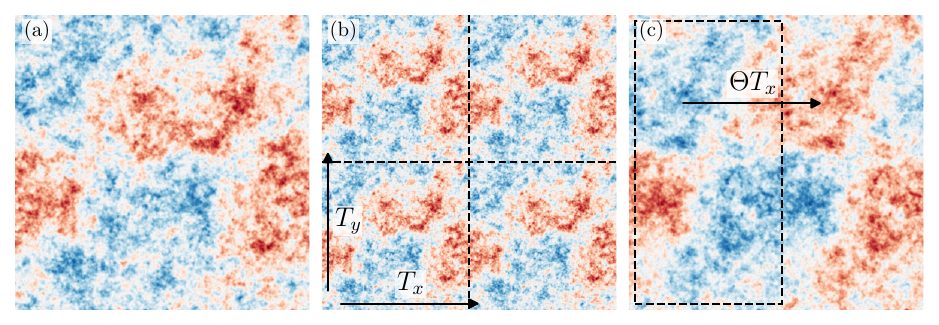}
  \caption{
    Disordered supercells with various symmetries at length scales larger than the localization length.
    (a) Disordered sample without any symmetry.
    (b) The disorder realization in (a) tiled with supercell translation symmetry.
    (c) A disordered supercell with magnetic translation symmetry.
  }
  \label{fig:timereversalsymmetry}
\end{figure}

\comment{Using a supercell is a standard technique for analyzing strong TIs.}
Before turning to statistical phases, we consider the problem of characterizing the topology of disordered strong topological phases.
As early as 1992, numerical work computed the Chern number of a supercell to probe the topology of disordered systems~\cite{Huo1992}.
While the supercell is mentioned only in passing, it represents a nontrivial identification.
Instead of a fully disordered system, schematically represented in Fig.~\ref{fig:timereversalsymmetry}(a), one considers a system with large but finite translational symmetries $T_x$ and $T_y$ [Fig.~\ref{fig:timereversalsymmetry}(b)].
Because topological properties are local, the topological invariant of the supercell converges to that of the disordered system as long as the supercell size $L$ exceeds the localization length at the Fermi level, $\xi$.

\comment{A supercell also works for weak TI as long as it is odd-sized.}
Reference~\cite{Ringel2012} proved the robustness of the surface states of weak topological insulators---the oldest known statistical topological phase---by considering the layer limit of these systems.
The authors argue that each layer carries a single pair of helical states and that an odd number of such pairs cannot be localized due to Kramers' theorem.
Because properties of a macroscopically large system are insensitive to the exact number of layers, the entire surface remains delocalized.
As a constructive application of this argument, the topological invariant of a large odd-sized supercell probes the bulk topology of a disordered weak topological insulator~\cite{Sbierski2014}.
Therefore, as in the strong topological insulator case, we enforce a translation symmetry, albeit with a more complex rule: an even-sized supercell always yields a trivial invariant.

\comment{We consider a generalization of the supercell procedure.}
The supercell procedure captures the topology of both strong and weak topological insulators.
For weak topological insulators, however, this holds only when the period is odd, which suggests a more general principle.
To formulate it, we consider a disordered ensemble specified by a probability distribution $P(H)$ over Hamiltonians $H$.
The ensemble has average symmetries $\mathcal{G}_a$ such that $P(H) = P(U_a H U_a^{-1})$ for $U_a \in \mathcal{G}_a$.
It also has an exact symmetry group $\mathcal{G}_e$ with $H = U_e H U_e^{-1}$ for all $H$ in the ensemble and $U_e \in \mathcal{G}_e$.
For any symmetry group, a fundamental domain is a subset that contains exactly one symmetry-inequivalent copy of each point; applying the symmetry operations to it tiles the full space.
We propose the following procedure to construct a new ensemble.
\begin{enumerate}
\item Select a suitable target exact symmetry group containing both $\mathcal{G}_e$ and a subgroup of the ensemble symmetry $\mathcal{G}_s \subset \mathcal{G}_a$.
In the case of the weak TI, $\mathcal{G}_s$ is the translation symmetry group of an odd-sized supercell.
\item For each Hamiltonian in the original ensemble, select all of its elements in the fundamental domain of the target symmetry group.
\item Tile the full space using the orbit of the selected fundamental domain.
\end{enumerate}
This procedure uses information only from a single fundamental domain of $\mathcal{G}_s$, so it discards some information about the original ensemble.
However, as long as the elements of $\mathcal{G}_s$ relate only points separated by more than $\xi$, the resulting ensemble is locally indistinguishable from the original.
The resulting ensemble also has a larger exact symmetry group $\mathcal{G}_e \times \mathcal{G}_s$, which is why we call it a \emph{symmetric approximant}.
Because this ensemble has an exact symmetry, its elements have well-defined topological invariants.
We shall see that in many cases it is possible to choose $\mathcal{G}_s$ so that the symmetric approximant captures the topology of the original ensemble.

\subsection{Mapping of average time-reversal onto magnetic translation}
\comment{The only possible embedding of an ensemble TRS is magnetic translation.}
We now apply the symmetric approximant to an open problem of computing the topological invariant of a 3D TI protected by average time-reversal symmetry (TRS) $\Theta$~\cite{Fulga2014}.
Reference~\cite{Fulga2014} considered adiabatically removing the TRS-breaking disorder.
However, this argument does not by itself provide an explicit numerical procedure for computing the invariant.
At first glance, there is no good choice of a target symmetry group $\mathcal{G}_s$.
Because $\Theta$ is a local symmetry, it is impossible to approximate an ensemble that breaks TRS locally with one that respects it exactly everywhere.
Likewise, the mapping requires that $\mathcal{G}_s$ relates only distant points.
Considering $\mathcal{G}_s$ generated by pure translation is insufficient because systematically breaking time-reversal symmetry trivializes the topological classification.
One final, previously overlooked, option is to consider a symmetry group generated by a magnetic translation $\Theta T_x$, where $T_x$ is a translation comparable to the supercell size and larger than the localization length, as shown in Fig.~\ref{fig:timereversalsymmetry}(c).
While $\Theta T_x$ shares some properties with $\Theta$, such as requiring the average magnetization to vanish, it is a nonlocal symmetry that preserves the local properties of the original ensemble.
Topological systems with magnetic translations are known as antiferromagnetic TIs~\cite{Mong2010}.

\comment{We show that mapping instead to an ensemble with magnetic-translation symmetry captures the topology of a 3D topological insulator with average TRS.}
This choice of symmetry group yields a symmetric approximant whose topology is equivalent to that of the original ensemble.
Because $(\Theta T_x)^2 = -\exp{i k_x}$ when applied to a Bloch state with momentum $k_x$, the topological invariant of an antiferromagnetic TI is~\cite{Mong2010}:
\begin{equation}
  Q_\textrm{AFTI} = Q_\textrm{QSHE}(k_x=0),
\end{equation}
where $Q_\textrm{QSHE}$ is the topological invariant of the quantum spin Hall phase.
The phase with average TRS is adiabatically connected to a time-reversal-symmetric phase with the same topology---trivial to trivial and topological to topological~\cite{Fulga2014}.
Because the approximant ensemble is locally indistinguishable from the original ensemble, the same adiabatic removal of TRS-breaking disorder that avoids delocalization in the original ensemble is guaranteed to work for the symmetric approximant.
Adiabatically removing the TRS-breaking disorder from the symmetric approximant without closing the bulk gap yields a time-reversal-symmetric system with a topological invariant~\cite{Fu2007}:
\begin{equation}
  Q_\textrm{3DTI} = Q_\textrm{QSHE}(k_x=0)Q_\textrm{QSHE}(k_x=\pi) = Q_\textrm{AFTI},
\end{equation}
where for the second equality we use that, once we remove the TRS-breaking disorder, the Brillouin zone folds and ensures that $k_x=\pi$ corresponds to a trivial quantum spin Hall invariant.
Therefore, both the original ensemble with average TRS and the symmetric approximant with magnetic-translation symmetry are adiabatically connected to the same clean phase.
We conclude that both ensembles share the same topological invariant.

\section{Local indistinguishability} \label{sec:indistinguishability}

\comment{Local indistinguishability ensures meaningful transitions.}
A key requirement of the symmetric approximant formalism is local indistinguishability between the approximant ensemble and the original ensemble under study.
Local indistinguishability ensures that any delocalization transition in the approximant corresponds to a delocalization transition in the original system.
In this section, we contrast two mappings: one that violates local indistinguishability and one that preserves it.
We show that violating local indistinguishability introduces phase transitions absent from the original disordered system.

\comment{Disorder can simplify classifications.}
We first apply the formalism to a disordered 2D second-order topological insulator in class $\textrm{AIII}^{\mathcal{M}_+}$.
This symmetry class consists of disordered Hamiltonians that respect chiral symmetry $\mathcal{C}$ and are drawn from an ensemble with an average mirror symmetry $\mathcal{M}$ that commutes with $\mathcal{C}$.
In the clean limit where $\mathcal{M}_y$ is an exact rather than average symmetry, the class has a $\ZZ$ classification of strong phases~\cite{geier2018}.
In the disordered case, where $\mathcal{M}_y$ is an average symmetry, the classification reduces to $\ZZ_2$~\cite{geier2018, Chaou2025}.

\comment{We study a layer-construction model that captures classification simplification.}
To apply the formalism, we first consider a clean mirror-symmetric system built from a layer-construction of higher-order topological insulators~\cite{isobe2015,fulga2016,huang2017,khalaf2018,trifunovic2019}.
The model~\cite{Chaou2025} has two layers per unit cell, and each layer consists of a pair of SSH wires aligned along the $x$-direction.
The Bloch Hamiltonian is
\begin{equation}
  H = \begin{pmatrix}
    h_+ + \sigma_2 \tau_1           & \delta \sigma_1 \tau_3 \sin k_y \\
    \delta \sigma_1 \tau_3 \sin k_y & h_- + \sigma_2(\tau_1 \cos k_y + \tau_2 \sin k_y )\end{pmatrix}.
\end{equation}
Here $h_{\pm} = \sigma_2 \tau_0 (M + \cos k_x) \pm \sigma_1 \tau_0 \sin k_x$.
The Pauli matrices $\sigma_i$ act in sublattice space, and $\tau_i$ act in the two-wire degree of freedom within each layer.
The $k_y$ dependence of the bottom-right block of $H$ arises because inter-wire couplings span unit cells.
Here $\delta$ sets the interlayer coupling, and $M$ is a mass parameter that drives the topological transition.
We consider the layered model near the 1D limit, achieved by choosing a small $\delta$, so that disorder can more easily localize the system.
This model preserves chiral symmetry $\mathcal{C} = \sigma_3$ and mirror symmetry $\mathcal{M}_y = \tau_1$.

\comment{Disorder simplifies the classification by blurring the distinction between two topological phases.}
In the clean limit, the model hosts three distinct phases.
It has a trivial phase for $|M|>2$ and two higher-order topological insulator (HOTI) phases with bulk invariants $Q_\textrm{bulk}=\pm1$ for $-2 < M < 0$ and $0 < M < 2$.
The topological phases are second-order HOTIs with corner modes at mirror-symmetric corners.
The phases are distinguished by the corner mode's eigenvalue under the product of chiral and mirror symmetries $\mathcal{C}\mathcal{M}_y$.
Adding disorder blurs this distinction and renders the two phases topologically equivalent, resulting in a single topological phase for $-2<M<2$ and $\alpha>0$~\cite{geier2018, Chaou2025}.
We simulate this topological phase by introducing intralayer disorder through random onsite and hopping terms that preserve chiral symmetry while breaking mirror symmetry.
Because one of the layers crosses the unit-cell boundary, part of this intralayer disorder appears as hopping terms in the Bloch Hamiltonian, even though the disorder remains local within each layer.
The disorder coefficients are drawn independently from a zero-mean Gaussian distribution, and the parameter $\alpha$ sets the overall disorder strength, so mirror symmetry is preserved only on average over the ensemble.

\begin{figure}[bth]
  \centering
  \includegraphics[width=\columnwidth]{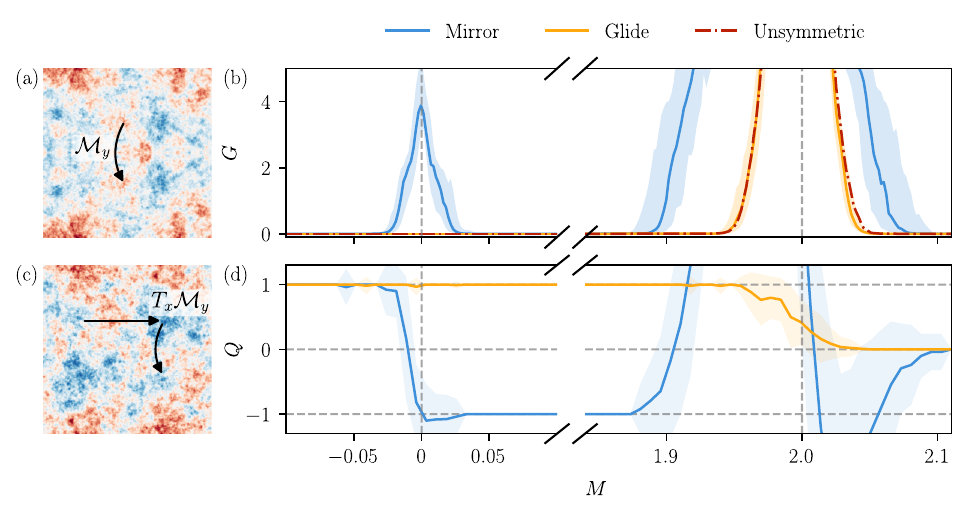}
  \caption{
    Invariant and conductances in mirror- and glide-symmetric systems.
    (a) Disorder sample with mirror symmetry.
    (b) The median conductance with the 16th--84th percentile band of a square with mirror- and glide-symmetric disorder and of a square with disorder that does not respect any symmetry.
    (c) Disorder sample with glide symmetry.
    (d) The mean of the topological invariant with $\pm \sigma$ for a system with mirror and glide symmetry.
    For the computation in (b/d) we use $\delta=0.1$, $\alpha = 0.5$, $L_{x,y}=200$ and 50 seeds.
  }
  \label{fig:mirrorsymmetry}
\end{figure}

\comment{We map onto an ensemble with exact global mirror symmetry in order to compute a scattering invariant.}
To show that local indistinguishability is crucial, we first attempt to compute a quantized invariant by enforcing an exact global mirror symmetry [depicted in Fig.~\ref{fig:mirrorsymmetry}(a)].
A scattering invariant~\cite{Zijderveld2025} becomes a natural choice\footnote{A Hamiltonian invariant could also be used; we choose a scattering invariant for simplicity.} for this purpose, because it is defined from the reflection matrix and can thus be computed in a finite disordered geometry.
We compute the reflection matrix $r$ for a square sample with mirror-symmetric disorder, periodic boundary conditions, and leads attached in the direction parallel to the mirror axis.
The reflection matrix $r$ is constrained by both chiral and mirror symmetries:
\begin{equation}
\begin{aligned}
  r = V_C r^{\dagger} Q_C^{\dagger}; \quad r = Q_{\mathcal{M}} r V_{\mathcal{M}}^{\dagger}.
\end{aligned}
\end{equation}
Here $V_\mathcal{O}$ and $Q_\mathcal{O}$ are unitary matrices that describe how incoming and outgoing lead wavefunctions transform under the symmetry operator $\mathcal{O}$.
We transform to a basis in which both symmetry operators are diagonal.
In this basis, $r$ is Hermitian and block-diagonal in the mirror eigenvalues.
The topological invariant is half the signature of the reflection matrix in one of the blocks, $\mathcal{Q} = \frac{1}{2} \operatorname{sig}(r_+) = -\frac{1}{2} \operatorname{sig}(r_-)$.

\comment{Exact global mirror symmetry introduces a transmitting mode absent in the unsymmetrized system.}
Although most of the phase diagram is captured by the parity of the invariant, $\mathcal{Q} \textrm{ mod } 2$, there is an issue apparent from Fig.~\ref{fig:mirrorsymmetry}(b): mirror-symmetric samples have a delocalization transition at $M=0$, while unsymmetrized samples do not.
This delocalization transition occurs because exact global mirror symmetry assigns a $\mathcal{C}\mathcal{M}_y$ parity to the corner modes.
This parity keeps the two topological phases distinct even with disorder and retains a $\ZZ$ classification rather than the reduced $\ZZ_2$ classification.
Interpolating between these two phases requires the bulk gap to close at $M=0$.
The gap closing occurs along the mirror axis because that axis is the only region locally different from an unsymmetrized sample.
Therefore, a perfectly transmitting 1D delocalized mode appears along the mirror axis at $M=0$.

\comment{The failure of the invariant is due to the lack of local indistinguishability, so to recover local indistinguishability we instead map average mirror symmetry to exact glide symmetry.}
The failure of the invariant to capture the simplified classification stems from introducing an \emph{exact} local mirror symmetry along the global mirror axis.
This introduction makes the new system locally \emph{distinguishable} from the original disordered system, so their phase transitions need not coincide.
To recover indistinguishability, we instead map the ensemble with average mirror symmetry $\mathcal{M}$ to an ensemble with exact glide symmetry $\mathcal{M}_y T_x$ with $T_x$ a large distance translation [see Fig.~\ref{fig:mirrorsymmetry}(c)].
Glide-symmetric systems with chiral symmetry in 2D have a $\ZZ_2$ classification~\cite{Shiozaki2016}, matching our target.

\comment{Diagnosing the phase of the approximant with a Hamiltonian invariant.}
To compute the invariant of the glide-symmetric approximant, we impose twisted boundary conditions along both spatial directions and evaluate the following Hamiltonian invariant of Ref.~\cite{Shiozaki2017}:
\begin{equation}
  \begin{aligned}
    \mathcal{Q} & = \operatorname{sign}
    \left[
      \frac{\det h_+(\pi,0)}{\det h_+(\pi,\pi)}
      \frac{\exp \left\{\tfrac{1}{2} \int_{-\pi}^{\pi} d{\phi_x} \ln \det h_+(\phi_x, 0) \right\} }{\exp \left\{ \tfrac{1}{2} \int_{-\pi}^{\pi} d{\phi_x} \ln \det h_+(\phi_x, \pi) \right\}  }
      \exp \left\{ \frac{1}{2} \int_{0}^{\pi} d {\phi_y} \ln \det h(\pi, \phi_y) \right\}
    \right].
  \end{aligned}
  \label{eq:axion-invariant}
\end{equation}
Here $\phi_{x,y}$ denote the boundary-condition twists along $x$ and $y$.
The terms $h_{\pm}(\phi_x, \phi_y)$ are the symmetry sectors of the off-diagonal blocks $h$ of the Hamiltonian in the glide basis:
\begin{equation}
  \mathcal{M}_y T_x =
  \begin{pmatrix}
    \sigma_0 \tau_1 & 0 \\
    0 & \sigma_0 \tau_1 e^{i \phi_x}
  \end{pmatrix}.
\end{equation}
This yields a quantized $\ZZ_2$ invariant that we compute for each disorder realization in the glide-symmetric ensemble.
For numerical efficiency, we use the sparse determinant function implemented by the MUMPS library~\cite{MUMPS2001, MUMPS2019}.
Figure~\ref{fig:mirrorsymmetry}(d) shows that, unlike the mirror-symmetric case, this invariant correctly captures the two phases of the system with average mirror symmetry.
There is no transition at $M=0$, and there is a transition at $M=2$.
This demonstrates that the symmetric approximant with exact glide symmetry $G$ diagnoses the topological phases of systems with average mirror symmetry $\mathcal{M}$.

\section{Intrinsic statistical topological phases} \label{sec:ISTI}

\comment{We apply the approximant symmetry formalism to a 2D ISTI in class D with inversion symmetry.}
We next apply the formalism to an ``intrinsic'' statistical topological insulator (ISTI), where disorder enables a topological phase with no clean counterpart~\cite{Chen2025}.
Such phases are difficult to characterize with conventional invariants because disorder averaging restores average symmetries to exact ones, making effective-Hamiltonian approaches ineffective~\cite{li_topological_2020,varjas_topological_2019,spring_isotropic_2023,marsal_obstructed_2023}.
We consider inversion-symmetric spinless superconductors in $2D$ where particle-hole symmetry $\mathcal{P}$ and $\mathcal{I}$ commute, \emph{i.e.} class $\textrm{D}^{\mathcal{I}_+}$.
This symmetry class cannot host a higher-order topological phase if it respects exact inversion $\mathcal{I}$ and $\mathcal{P}$~\cite{geier2018,trifunovic2019,Chaou2025}, because inversion would force an unpaired Majorana in each inversion-parity sector.
This restriction, however, no longer applies when inversion is preserved only on average.

\comment{We build a clean parent Hamiltonian and then add disorder to obtain an ISTI.}
To construct a disorder-induced phase, we start from a clean Hamiltonian built from a layer construction in which layers $a$ and $b$ are coupled by a hopping with strength $\delta$:
\begin{equation}
        H(\vk) = \begin{pmatrix}
        H_a(\vk) + M \tau_0\sigma_2  & \delta \sin k_y \tau_3 \sigma_0\\
        \delta \sin k_y \tau_3 \sigma_0 & H_b(\vk) + M \tau_0\sigma_2
        \end{pmatrix}.
\end{equation}
Each layer is made of two coupled wires that extend in the $x$ direction:
\begin{equation}
  \begin{aligned}
    H_a(\vk) &= (1 - \cos k_x) \tau_1\sigma_2 + \sin k_x \tau_2\sigma_2 \\
    H_b(\vk) &=
    (\cos k_y \tau_1+ \sin k_y \tau_2)\sigma_2-\cos (k_x-k_y) \tau_1\sigma_2 + \sin (k_x-k_y) \tau_2\sigma_2.
  \end{aligned}
\end{equation}
The Pauli matrices $\tau_i$ act on the two-wire degree of freedom in each layer, and $\sigma_i$ act on the orbital degree of freedom of each wire.
The Hamiltonian respects $\mathcal{I} = \sigma_1$ and $\mathcal{P} = \mathcal{K}$.
To obtain the intrinsic statistical topological phase, we add random onsite and hopping terms chosen to preserve particle-hole symmetry while breaking inversion in each realization.
Their coefficients are drawn independently from a zero-mean Gaussian distribution, and the parameter $\alpha$ sets the disorder strength, such that inversion symmetry is restored only on average over the ensemble.

\comment{To diagnose a topological phase, we find a correct mapping into an ensemble with exact $\mathcal{I}$ symmetry by considering local indistinguishability.}
The first step in applying the symmetric approximant formalism is to find a suitable exact symmetry to map into.
Unlike the previous examples, we consider a symmetric approximant without translation symmetry.
Instead, we map the system onto a Hamiltonian with exact \emph{global} $\mathcal{I}$ symmetry but without any translation symmetry.
This mapping leaves the $\mathcal{I}$-center point of the sample mapped onto itself, so the center is $\mathcal{I}$-invariant and distinct from the rest of the system.
Enforcing local indistinguishability therefore restricts us to geometries with a hole at the center.
This hole introduces an inner boundary that also supports a pair of zero-energy Majorana corner modes.

\begin{figure}[bth]
  \centering
  \includegraphics[width=\columnwidth]{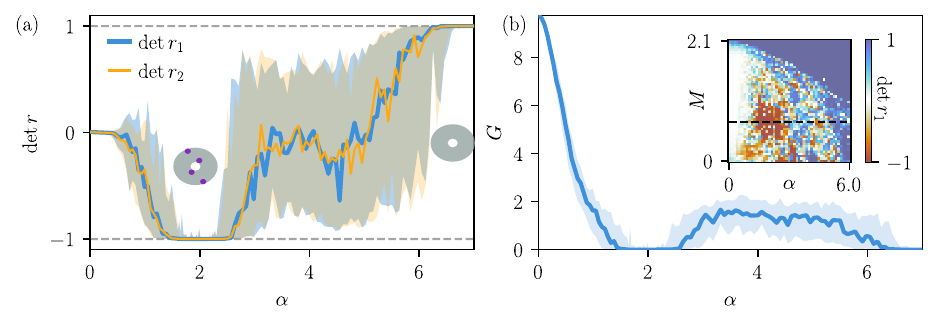}
  \caption{
    Invariant and conductance in a rotationally symmetric square with a hole.
    (a) Median determinants with 16th--84th percentile bands of the $\mathcal{I}$-symmetric blocks of the reflection matrix $r$ for a square system with a hole.
    The insets show a diagram of a Corbino disk with and without topological edge states.
    (b) The median conductance with the 16th--84th percentile bands of the same system.
    For this computation, we use $\delta=0.1$, $M = 0.7$, $L_{x,y}=300$, $R_{\mathrm{hole}}=1$, and 50 seeds.
    The inset shows the determinant of one of the $\mathcal{I}$-symmetric blocks as a function of both $\alpha$ and $M$ in a system with $L_{x,y}=200$, $R_{\mathrm{hole}}=1$, $\delta=0.1$ and for a single seed.
  }
  \label{fig:isti}
\end{figure}

\comment{The corresponding scattering invariant correctly characterizes the topology of the system.}
Such a finite setup naturally accommodates the use of a scattering invariant~\cite{Zijderveld2025}.
We attach leads to the inner and outer boundaries of the sample and compute the reflection matrix from the inner surface at the Fermi level.
We then constrain $r$ by both $\mathcal{I}$ and $\mathcal{P}$ symmetries:
\begin{equation}
    r = Q_{\mathcal{I}} r V_{\mathcal{I}}^{\dagger}; \quad r = V_{\mathcal{P}} r^{*} Q_{\mathcal{P}}^{\dagger}.
\end{equation}
Here $V_{\mathcal{O}}$ and $Q_{\mathcal{O}}$ are the unitary matrices that describe how the incoming and outgoing wavefunctions in the leads transform under the symmetry operator $\mathcal{O}$.
In the basis where $\mathcal P$ constrains $r$ to be real, the scattering invariant is
\begin{equation}
    \mathcal{Q}_{\pm} = \operatorname{sign}[\det r_{\pm}] = \pm 1.
\end{equation}
Here $r_\sigma$ is the reflection matrix in the $\mathcal{I}$-parity sector $\sigma$.
When the invariant is $\mathcal{Q}_\sigma = -1$, it indicates the presence of a Majorana zero mode with parity $\sigma$.
These two invariants are related because $\mathcal{I}$ forces both surfaces to host an even number of Majorana corner modes.
Therefore, $\det r = +1$ and $\mathcal Q_+ = \mathcal Q_-$.
The higher-order topological superconductor, which hosts an odd number of Majorana pairs, is thus identified by either $\mathcal Q_\pm = -1$.
Figure~\ref{fig:isti} shows the result.
A topological insulating region arises at finite disorder strength $\alpha$ but is absent both in the clean limit $\alpha=0$ and in the Anderson-insulating regime at strong disorder, $\alpha \gg 6$.
Panel (a) shows a quantized invariant for the two insulating regions shown in panel (b), where the conductance $G$ between the inner and outer leads vanishes.

\section{General construction and limitations} \label{sec:general}

\begin{figure}[bth]
  \centering
  \includegraphics[width=\columnwidth]{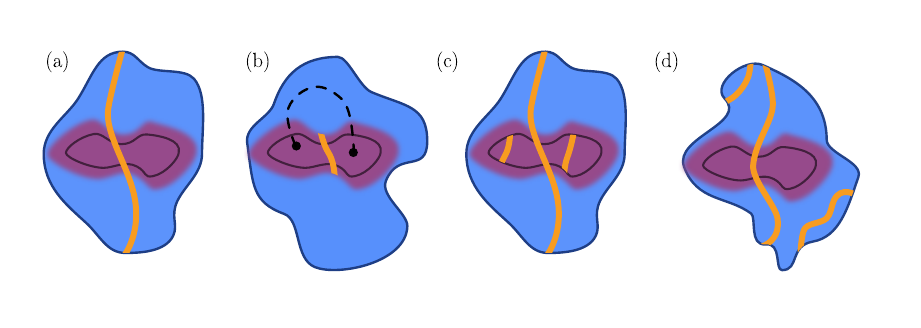}
  \caption{
    Possible relations between the topological classifications of the ensemble with an exact symmetry (purple region with a border), ensemble with an average symmetry (blurry purple region), and the symmetry class of the approximant (blue region that encloses the purple region).
    The orange lines show delocalization transitions.
    (a) The phase transition of the approximant symmetry class probes the phase transition of the disordered ensemble (Sec.~\ref{sec:magnetictranslation}).
    (b) The approximant ensemble admits a continuous deformation between topologically distinct phases of the original ensemble (dashed path).
    (c) The classification of the ensemble with the average symmetry is reduced compared to the exact one, and is captured by the approximant ensemble (Sec.~\ref{sec:indistinguishability}).
    (d) Some of the phase transitions of the symmetry class of the approximant do not correspond to those of the original ensemble.
  }
  \label{fig:blobs}
\end{figure}

\comment{We assess the generality of the approximant by comparing the classifications.}
To assess the generality of the symmetric approximant formalism, we compare the topological classifications of the original ensemble with those of the candidate approximant symmetry classes.
Every ensemble symmetry class allows multiple choices of the approximant symmetry, some of which capture the topological phases [Fig.~\ref{fig:blobs}(a)].
Other approximant choices have simpler topological classifications that capture either some of the target topological transitions or none at all [Fig.~\ref{fig:blobs}(b)].

\comment{The formalism avoids false positives, although this is insufficient to guarantee correctness if classifications coincide.}
The requirement of local indistinguishability is both stringent and powerful.
The example of Sec.~\ref{sec:indistinguishability} demonstrated that violating local indistinguishability introduces spurious phase transitions [Fig.~\ref{fig:blobs}(c)].
On the other hand, preserving local indistinguishability guarantees the appearance of delocalized states in the original ensemble whenever they appear in the approximant.
However, this does not guarantee that all topological transitions of the approximant's symmetry class correspond to those of the original ensemble.
The lower symmetry of the approximant symmetry class may yield additional topological phases [Fig.~\ref{fig:blobs}(d)].

\comment{STIs protected by sufficiently simple symmetries have a single candidate approximant.}
Focusing on the STIs protected by order-two average symmetries, \emph{i.e.} symmetries that square to the identity, we observe that only one or two candidate approximant symmetry classes preserve local indistinguishability, as shown in Table~\ref{tab:mappings}.
In particular, to guarantee that the image of every point is far from the original point, we either combine the additional symmetry with a translation or consider a pure (global) rotation symmetry and exclude the rotation center.
The symmetry classes protected by either an onsite average symmetry or average translation require a product of that symmetry with a translation or an odd-period supercell translation, respectively (Sec.~\ref{sec:magnetictranslation}).
These phases have a $\ZZ_2$ classification~\cite{Fulga2014} that is adiabatically connected to a clean invariant, and this guarantees the correctness of the approximant.

\begin{table}[h]
  \centering
  \begin{tabular}{c|c|c}
    Average symmetry & Mapping choice           & Failures        \\
    \hline
    Translation             & Odd-period translation   &                 \\
    Onsite                  & Onsite times translation &                 \\
    Mirror in 2D            & Glide                    &                 \\
    Mirror in 3D            & Glide                    & Critical hinges \\
    Rotation in 3D          & Screw or Rotation        & Third order     \\
    Inversion in 2D or 3D   & Inversion                &                 \\
  \end{tabular}
  \caption{Summary of mappings for statistical higher-order topological insulators.}
  \label{tab:mappings}
\end{table}

\comment{Changing reflection to glide only works partially.}
Reflection symmetry maps only to glide symmetry (Sec.~\ref{sec:indistinguishability}).
We compare the classifications of ensembles with average mirror symmetry~\cite{Chaou2025} with those of ensembles with exact glide symmetry~\cite{Shiozaki2017} (see App.~\ref{app:classifications_mirror}).
In 2D, glide symmetry has a larger or equal classification in all symmetry classes.
In 3D, the glide classification is smaller, for example, in the symmetry class $\textrm{D}^{\mathcal{M}_+}$.
We attribute this difference to the inability of the symmetric approximant to capture phases with a critical hinge.

\comment{Because rotations only leave the axis invariant, mapping into the exact symmetry class is sufficient.}
Unlike reflection, an inversion paired with a translation stays an inversion (with a shifted center), so it is the only choice.
While an inversion center breaks local indistinguishability, it does not introduce a bulk delocalization because of its zero-dimensional nature.
Because the classifying groups of all 2D inversion-symmetric ensembles with average (local) inversion symmetry are no larger than those with exact (global) inversion, the scattering invariant of Ref.~\cite{Zijderveld2025} correctly captures their topology.
The same reasoning applies to 3D inversion-symmetric ensembles~\cite{Song2021,Chaou2025}, although the scattering invariant of third-order topological phases and the classification of inversion-protected ISTIs are unknown.
Finally, 3D rotation-symmetric ensembles with average rotation symmetry map either to ensembles with exact rotation or screw symmetry.
The former requires a sample with a hole at the rotation axis, similar to Ref.~\cite{Zijderveld2025}.
Both approaches work for first- and second-order phases, but they are insufficient to capture third-order phases.
This happens because the classification of third-order phases with average rotation symmetry reduces compared to those with exact rotation symmetry~\cite{Chaou2025}.
At the same time, screw-symmetric samples cannot host third-order phases because decorating their surface with a screw-symmetric spiral removes the zero modes.
This reasoning matches the differences of the corresponding classifications listed in App.~\ref{app:classifications_rotation}.

\section{Conclusion} \label{sec:conclusion}

\comment{We demonstrated that in many cases the symmetric approximant formalism works.}
We demonstrated that in many cases the topological invariants of statistical topological phases are captured by their symmetric approximants.
The approximant works by enforcing an exact symmetry in place of an average one while staying locally indistinguishable from the original ensemble.
The additional exact symmetry changes the topology from being a property of the entire ensemble to being a property of every disorder realization.
The approximant approach applies to multiple types of statistical topological phases:
\begin{itemize}
  \item statistical phases with critical boundaries (Sec.~\ref{sec:magnetictranslation}),
  \item average spatial symmetries that reduce the topological classification (Sec.~\ref{sec:indistinguishability}),
  \item intrinsic STIs that have no clean counterpart (Sec.~\ref{sec:ISTI}).
\end{itemize}
Our construction fails to capture third-order phases and critical hinge boundaries.
For that reason, we are unable to answer affirmatively whether all statistical topological phases have a conventional topological invariant.
However, we consider the partial success of the formalism a strong indication that such invariants exist.

\comment{The extensions of our work range from straightforward to involved and highly nontrivial.}
While we treated only uncorrelated disorder, extending the procedure to short-range correlated disorder differs only in how the approximant ensemble is generated.
Because the symmetric approximant typically has a nonsymmorphic symmetry, extending the formalism to more complex symmetry groups requires further development of the theory of topological insulators in nonsymmorphic crystals.
It remains unclear to what extent the symmetric approximant formalism applies to statistical symmetry-protected interacting phases that share some phenomenology~\cite{Ma2023, Ma2025}.

\section*{Acknowledgements}
We thank I.~C.~Fulga and P.~W.~Brouwer for useful discussions.
I.~A.~D.~acknowledges financial support from the Netherlands Organization for Scientific Research (NWO/OCW) as part of the Frontiers of Nanoscience program.
A.~Y.~C.~acknowledges financial support from the Deutsche Forschungsgemeinschaft (DFG, German Research Foundation), project number 277101999, CRC TR 183 (projects A03).
A.~Y.~C.~and I.~A.~D.~acknowledge financial support from the European Research Council (ERC) Consolidator grant under grant agreement number 101042707 TOPOMORPH.

\section*{AI disclosure}
We have used GitHub Copilot for typing completions and prompt-based code generation and modification.
We used ChatGPT to generate critical evaluations of text and language quality.
We used GitHub Copilot and OpenAI Codex to produce rule-based text rewriting for consistent style.
These tools were used with a range of models: GPT-4--GPT-5.2, GPT-5.4, GPT-5.5, and GPT-5.6 from OpenAI and Claude Sonnet and Opus versions 4 and 4.5 from Anthropic (only in Copilot).
All AI contributions were reviewed and edited by the authors, who take full responsibility for the content of the publication.

\section*{Data availability}
The code used to produce the reported results and the generated data are available on Zenodo~\cite{ZenodoCode}.
The numerical calculations and figures used Kwant, NumPy, SciPy, Matplotlib, and Qsymm~\cite{Groth_2014,2020NumPy-Array,2020SciPy-NMeth,Hunter:2007,Varjas_2018}.

\section*{Author contributions}
All authors contributed to defining the initial and the final project scope, brainstorming the approach, writing code and writing the manuscript.
R.~J.~Z.~led the coding and wrote an initial draft.
A.~A.~oversaw the project.

\bibliography{bibliography}

\appendix
\input{appendix_tables.tex}

\end{document}

%% file: appendix_tables.tex
\section{Mirror symmetric phases} \label{app:classifications_mirror}

For mirror symmetric phases in 2D we conjecture that the correct choice of mapping is to apply a half translation parallel to the mirror axis, as was done for the example in Sec.~\ref{sec:indistinguishability}.
This choice of mapping may capture all STIs protected by average mirror symmetry in 2D because the nonsymmorphic classification is at least as rich as the statistical classification, as shown in Table~\ref{tab:classifications_mirror}.
The nonsymmorphic classifications for each symmetry class are taken from Table VI of Ref.~\cite{Shiozaki2017}.
The statistical classifications are taken from Table IV of Ref.~\cite{Chaou2025}.
We note that the statistical classifications do not include intrinsic statistical topological phases.

\begin{figure}[bth]
  \centering
  \includegraphics[width=\columnwidth]{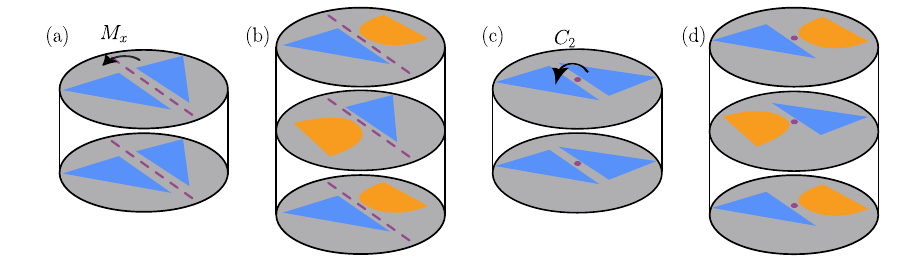}
  \caption{
    (a) shows a 3D system with mirror symmetry.
    (b) shows a 3D system with glide symmetry.
    (c) shows a 3D system with $C_2$ rotation symmetry.
    (d) shows a 3D system with screw symmetry.
  }
  \label{fig:cylinders}
\end{figure}

In 3D with average mirror symmetry, we consider a mapping choice where we translate along a direction in the mirror plane.
For example:
\begin{equation}
    (x, y, z) \rightarrow (-x, y, z+1/2).
\end{equation}
Figure~\ref{fig:cylinders}(a) and (b) show examples of systems with mirror and glide symmetry in 3D, respectively.
In Table~\ref{tab:classifications_mirror} we compare the classifications in Table VI of Ref.~\cite{Shiozaki2017} with the statistical classifications in Table VI of Ref.~\cite{Chaou2025}.
The comparison shows that a nonsymmorphic mapping only captures all phases for symmetry classes that do not host critical statistical hinge states as their boundary signature.
These symmetry classes have been highlighted in blue in Table~\ref{tab:classifications_mirror}.

\begin{center}
     \begin{table}
    \centering
    \begin{tabular}{l|cc|cc}
\hline\hline
\multirow{2}{*}{Class} & \multicolumn{2}{c|}{2D} & \multicolumn{2}{c}{3D} \\
\cline{2-5}
        & Statistical             & Nonsymmorphic          & Statistical   & Nonsymmorphic \\

\hline
    $\text{A}^\mathcal{M}$         & $-$                                                & $-$                                               & $\mathbb{Z}_2$                     & $\mathbb{Z}_2$                        \\
    $\text{AIII}^\mathcal{M_+}$    & $\mathbb{Z}_2$                                     & $\mathbb{Z}_2$                                    & $-$                                & $-$                                   \\
    $\text{A}^\mathcal{CM}$        & $\mathbb{Z}$                                       & $\mathbb{Z}$                                      & $-$                                & $-$                                   \\
    $\text{AIII}^\mathcal{M_-}$    & $-$                                                & $-$                                               & $\bl{\mathbb{Z} \times \mathbb{Z}_2}$ & $\mathbb{Z}$                         \\
    $\text{A}^\mathcal{T^+M}$      & $\mathbb{Z}$                                       & $\mathbb{Z}$                                      & $-$                                & $-$                                   \\
    $\text{AIII}^\mathcal{T^+M_+}$ & $\mathbb{Z}_2$                                     & $\mathbb{Z}_2$                                    & $\mathbb{Z}$                       & $\mathbb{Z}$                          \\
    $\text{A}^\mathcal{P^+M}$      & $-$                                                & $-$                                               & $\mathbb{Z}_2$                     & $\mathbb{Z}_2$                        \\
    $\text{AIII}^\mathcal{T^-M_-}$ & $-$                                                & $-$                                               & $\bl{\mathbb{Z}_2}$                   & $-$                                   \\
    $\text{A}^\mathcal{T^-M}$      & $2\mathbb{Z}$                                     & $\mathbb{Z}$                                      & $-$                                & $-$                                   \\
    $\text{AIII}^\mathcal{T^-M_+}$ & $-$                                                & $\mathbb{Z}_2$                                    & $2\mathbb{Z}$                      & $\mathbb{Z}$                         \\
    $\text{A}^\mathcal{P^-M}$      & $-$                                                & $-$                                               & $-$                                & $\mathbb{Z}_2$                        \\
    $\text{AIII}^\mathcal{T^+M_-}$ & $-$                                                & $-$                                               & $-$                                & $-$                                   \\
    $\text{AI}^\mathcal{M_+}$      & $-$                                                & $-$                                               & $-$                                & $-$                                   \\
    $\text{BDI}^\mathcal{M_{++}}$  & $\mathbb{Z}_2$                                     & $\mathbb{Z}_2$                                    & $-$                                & $-$                                   \\
    $\text{D}^\mathcal{M_{+}}$     & $\mathbb{Z}_2$                                     & $\mathbb{Z}_2$                                    & $\bl{\mathbb{Z}_4}$                   & $\mathbb{Z}_2$                     \\
    $\text{DIII}^\mathcal{M_{++}}$ & $\mathbb{Z}_2$                                     & $\mathbb{Z}_4$                                    & $\mathbb{Z}_2$                     & $\mathbb{Z}_2$                        \\
    $\text{AII}^\mathcal{M_+}$     & $-$                                                & $\mathbb{Z}_2$                                    & $\mathbb{Z}_2$                     & $\mathbb{Z}_4$                        \\
    $\text{CII}^\mathcal{M_{++}}$  & $\mathbb{Z}_2$                                     & $\mathbb{Z}_2$                                    & $-$                                & $\mathbb{Z}_2$                        \\
    $\text{C}^\mathcal{M_+}$       & $-$                                                & $-$                                               & $\mathbb{Z}_2$                     & $\mathbb{Z}_2$                        \\
    $\text{CI}^\mathcal{M_{++}}$   & $-$                                                & $-$                                               & $-$                                & $-$                                   \\
    $\text{AI}^\mathcal{CM_-}$     & $-$                                                & $-$                                               & $-$                                & $-$                                   \\
    $\text{BDI}^\mathcal{M_{+-}}$  & $-$                                                & $-$                                               & $-$                                & $-$                                   \\
    $\text{D}^\mathcal{CM_{+}}$    & $\mathbb{Z} \times \mathbb{Z}_2$                   & $\mathbb{Z} \times \mathbb{Z}_2$                  & $-$                                & $-$                                   \\
    $\text{DIII}^\mathcal{M_{-+}}$ & $\mathbb{Z}_2 \times \mathbb{Z}_2$                                    & $\mathbb{Z}_2 \times \mathbb{Z}_2$                 & $\bl{\mathbb{Z} \times \mathbb{Z}_4}$ & $\mathbb{Z} \times \mathbb{Z}_2$   \\
    $\text{AII}^\mathcal{CM_-}$    & $\mathbb{Z}_2$                                     & $\mathbb{Z}_2$                                    & $\mathbb{Z}_2 \times \mathbb{Z}_2$                      & $\mathbb{Z}_2 \times \mathbb{Z}_2$ \\
    $\text{CII}^\mathcal{M_{+-}}$  & $-$                                                & $-$                                               & $\bl{\mathbb{Z}_2 \times \mathbb{Z}_2}$                 & $\mathbb{Z}_2$                     \\
    $\text{C}^\mathcal{CM_+}$      & $2\mathbb{Z}$                                      & $2\mathbb{Z}$                                     & $-$                                & $-$                                   \\
    $\text{CI}^\mathcal{M_{-+}}$   & $-$                                                & $-$                                               & $2\mathbb{Z}$                      & $2\mathbb{Z}$                         \\
    $\text{AI}^\mathcal{M_-}$      & $-$                                                & $-$                                               & $-$                                & $-$                                   \\
    $\text{BDI}^\mathcal{M_{--}}$  & $-$                                                & $\mathbb{Z}_2$                                    & $-$                                & $-$                                   \\
    $\text{D}^\mathcal{M_-}$       & $-$                                                & $\mathbb{Z}_2$                                    & $-$                                & $\mathbb{Z}_2$                        \\
    $\text{DIII}^\mathcal{M_{--}}$ & $\mathbb{Z}_4$                                     & $\mathbb{Z}_4$                                    & $-$                                & $\mathbb{Z}_2$                        \\
    $\text{AII}^\mathcal{M_-}$     & $\mathbb{Z}_2$                                     & $\mathbb{Z}_2$                                    & $\mathbb{Z}_4$                     & $\mathbb{Z}_4$                        \\
    $\text{CII}^\mathcal{M_{--}}$  & $\mathbb{Z}_2$                                     & $\mathbb{Z}_2$                                    & $\mathbb{Z}_2$                     & $\mathbb{Z}_2$                        \\
    $\text{C}^\mathcal{M_-}$       & $-$                                                & $-$                                               & $\mathbb{Z}_2$                     & $\mathbb{Z}_2$                        \\
    $\text{CI}^\mathcal{M_{--}}$   & $-$                                                & $-$                                               & $-$                                & $-$                                   \\
    $\text{AI}^\mathcal{CM_+}$     & $-$                                                & $-$                                               & $-$                                & $-$                                   \\
    $\text{BDI}^\mathcal{M_{-+}}$  & $-$                                                & $-$                                               & $\bl{\mathbb{Z}_2}$                   & $-$                                \\
    $\text{D}^\mathcal{CM_-}$      & $\mathbb{Z}$                                      & $\mathbb{Z} \times \mathbb{Z}_2$                   & $-$                                & $-$                                   \\
    $\text{DIII}^\mathcal{M_{+-}}$ & $-$                                                & $\mathbb{Z}_2 \times \mathbb{Z}_2$                 & $\mathbb{Z}$                          & $\mathbb{Z} \times \mathbb{Z}_2$   \\
    $\text{AII}^\mathcal{CM_+}$    & $\mathbb{Z}_2$                                     & $\mathbb{Z}_2$                                    & $-$                                   & $\mathbb{Z}_2 \times \mathbb{Z}_2$ \\
    $\text{CII}^\mathcal{M_{-+}}$  & $-$                                                & $-$                                               & $\bl{\mathbb{Z}_4}$                   & $\mathbb{Z}_2$                     \\
    $\text{C}^\mathcal{CM_-}$      & $\mathbb{Z}$                                      & $2\mathbb{Z}$                                     & $-$                                & $-$                                   \\
    $\text{CI}^\mathcal{M_{+-}}$   & $-$                                                & $-$                                               & $\mathbb{Z}$                          & $2\mathbb{Z}$                      \\
    \hline\hline

    \end{tabular}

     \caption{Classification of statistical topological phases protected by average mirror symmetry and classification of order 2 nonsymmorphic symmetry classes in both 2D and 3D.}
    \label{tab:classifications_mirror}
\end{table}
\end{center}

\section{Rotation symmetric phases in 3D} \label{app:classifications_rotation}

For statistical topological phases protected by average rotation symmetry in 3D, one potentially suitable mapping choice is to consider the nonsymmorphic subgroup with screw symmetry.
Specifically, for a $C_2$ rotation in the $x,y$-plane, the screw transformation is:
\begin{equation}
    (x, y, z) \rightarrow (-x, -y, z + 1/2).
\end{equation}
We show examples of systems with rotation and screw symmetry in 3D in Fig.~\ref{fig:cylinders}(c) and (d), respectively.
In Table~\ref{tab:classifications_rotation} we compare the classifications in Table V of Ref.~\cite{Shiozaki2017} with the statistical classifications in Table VII of Ref.~\cite{Chaou2025}.
The subgroup sequences of the statistical classifications are also shown, because these show which symmetry classes have third-order edge modes present.
For an explanation of how to read the subgroup sequences, see Fig.~6 of Ref.~\cite{Chaou2025}.
The nonsymmorphic mapping choice may be appropriate in all symmetry classes, except for the symmetry classes that have third-order phases, which are shown in teal.

\begin{center}
    \begin{table}
    \centering
    \begin{tabular}{l|cc}
        \hline\hline
        Class ($d=3$)                  & Statistical                       & Nonsymmorphic      \\
        \hline
        $\text{A}^\mathcal{R}$         & $-$                                                                & $-$                                \\
        $\text{AIII}^\mathcal{R_+}$    & $\gr{0 \subseteq \ZZ_2 \subseteq \ZZ_2 \subseteq \ZZ\times\ZZ_2}$  & $\mathbb{Z}$                       \\
        $\text{A}^\mathcal{CR}$        & $0 \subseteq 0     \subseteq \ZZ_2 \subseteq \ZZ_2         $       & $\mathbb{Z}_2$                     \\
        $\text{AIII}^\mathcal{R_-}$    & $-$                                                                & $-$                                \\
        $\text{A}^\mathcal{T^+R}$      & $ 0 \subseteq 0     \subseteq \ZZ_2 \subseteq \ZZ_2         $      & $\mathbb{Z}_2$                     \\
        $\text{AIII}^\mathcal{T^+R_+}$ & $ \gr{0 \subseteq \ZZ_2 \subseteq \ZZ_2 \subseteq \ZZ_2}         $ & $-$                                \\
        $\text{A}^\mathcal{P^+R}$      & $-$                                                                & $-$                                \\
        $\text{AIII}^\mathcal{T^-R_-}$ & $ 0 \subseteq 0     \subseteq 0     \subseteq 2\ZZ          $      & $\mathbb{Z}$                       \\
        $\text{A}^\mathcal{T^-R}$      & $-$                                                                & $\mathbb{Z}_2$                     \\
        $\text{AIII}^\mathcal{T^-R_+}$ & $-$                                                                & $-$                                \\
        $\text{A}^\mathcal{P^-R}$      & $-$                                                                & $-$                                \\
        $\text{AIII}^\mathcal{T^+R_-}$ & $ 0 \subseteq 0     \subseteq 0     \subseteq \ZZ           $      & $\mathbb{Z}$                       \\
        $\text{AI}^\mathcal{R_+}$      & $-$                                                                & $-$                                \\
        $\text{BDI}^\mathcal{R_{++}}$  & $ \gr{0 \subseteq \ZZ_2 \subseteq \ZZ_2 \subseteq \ZZ_2}         $ & $-$                                \\
        $\text{D}^\mathcal{R_{+}}$     & $-$                                                                & $-$                                \\
        $\text{DIII}^\mathcal{R_{++}}$ & $ 0 \subseteq 0     \subseteq 0     \subseteq \ZZ           $      & $\mathbb{Z} \times \mathbb{Z}_2$   \\
        $\text{AII}^\mathcal{R_+}$     & $-$                                                                & $\mathbb{Z}_2 \times \mathbb{Z}_2$ \\
        $\text{CII}^\mathcal{R_{++}}$  & $ 0 \subseteq 0     \subseteq 0     \subseteq \ZZ_2         $      & $\mathbb{Z}_2$                     \\
        $\text{C}^\mathcal{R_+}$       & $-$                                                                & $-$                                \\
        $\text{CI}^\mathcal{R_{++}}$   & $ 0 \subseteq 0     \subseteq 0     \subseteq \ZZ  $               & $2\mathbb{Z}$                      \\
        $\text{AI}^\mathcal{CR_-}$     & $-$                                                                & $-$                                \\
        $\text{BDI}^\mathcal{R_{+-}}$  & $-$                                                                & $-$                                \\
        $\text{D}^\mathcal{CR_{+}}$    & $ \gr{0 \subseteq \ZZ_2 \subseteq \ZZ_4 \subseteq \ZZ_4}         $ & $\mathbb{Z}_2$                     \\
        $\text{DIII}^\mathcal{R_{-+}}$ & $ 0 \subseteq 0     \subseteq \ZZ_2 \subseteq \ZZ_2         $      & $\mathbb{Z}_2$                     \\
        $\text{AII}^\mathcal{CR_-}$    & $ 0 \subseteq 0     \subseteq \ZZ_2 \subseteq \ZZ_2         $      & $\mathbb{Z}_4$                     \\
        $\text{CII}^\mathcal{R_{+-}}$  & $-$                                                                & $\mathbb{Z}_2$                     \\
        $\text{C}^\mathcal{CR_+}$      & $ 0 \subseteq 0     \subseteq \ZZ_2 \subseteq \ZZ_2         $      & $\mathbb{Z}_2$                     \\
        $\text{CI}^\mathcal{R_{-+}}$   & $-$                                                                & $-$                                \\
        $\text{AI}^\mathcal{R_-}$      & $-$                                                                & $-$                                \\
        $\text{BDI}^\mathcal{R_{--}}$  & $-$                                                                & $-$                                \\
        $\text{D}^\mathcal{R_-}$       & $-$                                                                & $-$                                \\
        $\text{DIII}^\mathcal{R_{--}}$ & $ \gr{0 \subseteq \ZZ_2 \subseteq \ZZ_4 \subseteq \ZZ\times\ZZ_4}$ & $\mathbb{Z} \times \mathbb{Z}_2$   \\
        $\text{AII}^\mathcal{R_-}$     & $ 0 \subseteq 0     \subseteq \ZZ_2 \subseteq \ZZ_2 \times \ZZ_2       $      & $\mathbb{Z}_2 \times \mathbb{Z}_2$ \\
        $\text{CII}^\mathcal{R_{--}}$  & $ \gr{0 \subseteq \ZZ_2 \subseteq \ZZ_2 \subseteq \ZZ_2 \times \ZZ_2}       $ & $\mathbb{Z}_2$                     \\
        $\text{C}^\mathcal{R_-}$       & $-$                                                                & $-$                                \\
        $\text{CI}^\mathcal{R_{--}}$   & $ 0 \subseteq 0     \subseteq 0     \subseteq 2\ZZ          $      & $2\mathbb{Z}$                      \\
        $\text{AI}^\mathcal{CR_+}$     & $-$                                                                & $-$                                \\
        $\text{BDI}^\mathcal{R_{-+}}$  & $-$                                                                & $-$                                \\
        $\text{D}^\mathcal{CR_-}$      & $-$                                                                & $\mathbb{Z}_2$                     \\
        $\text{DIII}^\mathcal{R_{+-}}$ & $-$                                                                & $\mathbb{Z}_2$                     \\
        $\text{AII}^\mathcal{CR_+}$    & $ 0 \subseteq 0     \subseteq \ZZ_2 \subseteq \ZZ_4         $      & $\mathbb{Z}_4$                     \\
        $\text{CII}^\mathcal{R_{-+}}$  & $ 0 \subseteq 0     \subseteq 0     \subseteq \ZZ_2         $      & $\mathbb{Z}_2$                     \\
        $\text{C}^\mathcal{CR_-}$      & $ 0 \subseteq 0     \subseteq \ZZ_2 \subseteq \ZZ_2         $      & $\mathbb{Z}_2$                     \\
        $\text{CI}^\mathcal{R_{+-}}$   & $-$                                                                & $-$                                \\
        \hline\hline
    \end{tabular}
    \caption{Classification and subgroup sequences of statistical topological phases protected by average rotation symmetry in 3D and classification of nonsymmorphic symmetry classes in 3D.}
    \label{tab:classifications_rotation}
\end{table}
\end{center}